\documentclass{amsart}

\usepackage{graphicx}
\usepackage{color}
\newtheorem{theorem}{Theorem}
 \newtheorem{lem}{Lemma}

  \newtheorem{remark}{Remark}
\newtheorem{corollary}{Corollary}
\UseRawInputEncoding
\usepackage[russian, english]{babel}

\def\dfrac#1#2{\displaystyle{#1\over #2}}

\def\zo{\omega}
\def\zt{\theta}
\def\bv{{\bf v}}
\def\bV{{\bf V}}
\def\bp{{\bf p}}

\def\zr{\rho}

\def\Div{\mbox{div}\,}

\def\bB{{\bf B}}

\def\bE{{\bf E}}

\begin{document}

\markboth{Rozanova}{The Interplay of Regularizing Factors}

%
%

\title[Small perturbations in the Model of Upper Hybrid Oscillations]{
STUDY OF SMALL PERTURBATIONS OF A STATIONARY STATE IN A MODEL OF UPPER HYBRID PLASMA OSCILLATIONS
}

\author{Olga S. Rozanova}

\address{ Mathematics and Mechanics Department, Lomonosov Moscow State University, Leninskie Gory,
Moscow, 119991,
Russian Federation,
rozanova@mech.math.msu.su}



\maketitle


\begin{abstract}
 It is shown that a constant external magnetic field, generally speaking, is not able to prevent breaking (loss of smoothness) of relativistic plasma oscillations, even if they are arbitrarily small perturbations of the zero steady state. This result sharply differs from the non-relativistic case, for which it is possible to suppress the breaking  of oscillations at any initial deviations by increasing the intensity of the magnetic field \cite {RCharx}. Nevertheless, even in the relativistic case, there are subclasses of solutions corresponding to solutions that are globally smooth in time.
 \end{abstract}


\section{Introduction}	
The system of equations of hydrodynamics of "cold" plasma  \cite {ABR78}, \cite {GR75} 
  has the form
\begin{equation}
\label{base1}
\begin{array}{c}
\dfrac{\partial n }{\partial t} + \Div(n \bv)=0\,,\quad
\dfrac{\partial \bp }{\partial t} + \left( \bv \cdot \nabla \right) \bp
= e \, \left( \bE + \dfrac{1}{c} \left[\bv \times  \bB\right]\right),\vspace{0.5em}\\
\gamma = \sqrt{ 1+ \dfrac{|\bp|^2}{m^2c^2} }\,,\quad
\bv = \dfrac{\bp}{m \gamma}\,,\vspace{0.5em}\\
\dfrac1{c} \frac{\partial \bE }{\partial t} = - \dfrac{4 \pi}{c} e n \bv
 + {\rm rot}\, \bB\,,\quad
\dfrac1{c} \frac{\partial \bB }{\partial t}  =
 - {\rm rot}\, \bE\,, \quad \Div \bB=0\,,
\end{array}
\end{equation}
where
$ e, m $ are
charge and mass of the electron (here the electron charge has a negative sign: $ e <0 $),
$c$ is the speed of light;
$ n, \bp, \bv$ are the density, momentum and velocity of
electrons;
$\gamma$ is the  Lorentz factor;
$ \bE, \bB $ are vectors of electric and magnetic fields.
Plasma is considered a relativistic electron (fully ionized) liquid, neglecting recombination effects and ion motion.

The full system of equations \eqref {base1} is very difficult even for numerical simulation (monograph ~\cite {Ch_book} is devoted to these questions). It is well known that even relatively small initial perturbations can cause large-amplitude oscillations. Their evolution, as a rule, leads to the emergence of a strong singularity of the electron density ~ \cite {david72}, which is commonly called {\it breaking} of oscillations. For the rest of the solution components, the breaking process means the formation of  infinite  gradients.

Investigation of the dynamics of a plasma placed in an external magnetic field is a separate very complex issue. Analytical approaches, as a rule, are limited to the study of linearized models, for which one has to make assumptions about the smallness of the stationary state perturbations at all times. Even such linear waves are so complex that a special classification ~ \cite {ABR78}, \cite {GR75} is introduced for them. However, linear waves cannot help in studying the breaking phenomenon. Therefore, nonlinear models are of particular value, albeit greatly simplified in comparison with the original \eqref {base1} system, but retaining its important features and allowing to assess the possibility of delaying or completely eliminating the moment of breaking. It is they that are important for analyzing the acceleration of electrons in the wake wave of a powerful laser pulse~\cite{esarey09}.

One of such classical nonlinear models is the so-called {\it upper hybrid oscillation} model. It looks like it first appeared in ~ \cite {david72}, but has been repeatedly investigated in different contexts at the physical level of rigor (for example, \cite {Kar16}, \cite {Maity13}, \cite {Maity12}, \cite { Maity13_PRL}, \cite {verma} and literature cited there). It is assumed that the plasma is in a constant external magnetic field $ \bB = (0,0, B_0) $. The model describes the nonlinear dynamics of plasma oscillations propagating perpendicular to the external magnetic field. If we assume that the oscillations depend only on one spatial variable $ x $, then we obtain the following structure of the vectors of the momentum $ \bp $ and the electric field $\bE$~\cite{david72}:
$$
\bp(x,t) = (P_1(x,t),P_2(x,t),0), \quad  \bE(x,t) = (E_1(x,t),0,0).
$$
Condition of electrostaticity
\begin{equation*}
\label{rot0}
{\rm rot}\, \bE = 0,
\end{equation*}
holds automatically.

The system of upper hybrid oscillations has the form
\begin{equation}
\begin{array}{c}
\dfrac{\partial n }{\partial t} +
\dfrac{\partial }{\partial x}
\left(n\, V_1 \right)
=0,\quad
\dfrac{\partial P_1 }{\partial t} + V_1 \dfrac{\partial P_1}{\partial x}= e\left[ E_1 + \dfrac1{c}\, V_2 B_0\right],
\quad
{V_1} = \dfrac{P_1}{m \,\gamma}, \vspace{1 ex}\\
\dfrac{\partial P_2 }{\partial t} + V_1 \dfrac{\partial P_2}{\partial x}= - \dfrac{e}{c}\,V_1  B_0,
\quad {V_2} = \dfrac{P_2}{m \,\gamma}, \quad
\gamma = \sqrt{ 1+ \dfrac{P_1^2 + P_2^2}{m^2c^2}}\,,  \quad \vspace{1 ex}\\
\dfrac{\partial E_1 }{\partial t} = - 4 \,\pi \,e \,n\, V_1.
\end{array}
\label{3gl2}
\end{equation}
Note that it is not a direct consequence of  (\ref {base1}) system, but retains its basic features.
We introduce the dimensionless quantities
$$
\begin{array}{c}
\rho = k_p x, \quad \theta = \omega_p t, \quad
{\hat V_1} = \dfrac{V_1}{c}, \quad
{\hat P_1} = \dfrac{P_1}{m\,c}, \quad
{\hat V_2} = \dfrac{V_2}{c}, \quad
{\hat P_2} = \dfrac{P_2}{m\,c}, \vspace{1 ex}\\
{\hat E_1} = -\,\dfrac{e\,E_1}{m\,c\,\omega_p}, \quad
{\hat N} = \dfrac{n}{n_0}, \quad
{\hat B}_0 = -\,\dfrac{e\,B_0}{m\,c\,\omega_p},
\end{array}
$$
where $ \omega_p = \left(4 \pi e^2 n_0/m \right)^{1/2}$ is the plasma frequency, $n_0$ is  the unperturbed electronic
density, $k_p = \omega_p /c$.
In the new variables, system ~ (\ref {3gl2})  takes the form
\begin{equation}
\begin{array}{c}
\dfrac{\partial {\hat N} }{\partial \zt} +
\dfrac{\partial }{\partial \zr}
\left({\hat N}\, {\hat V}_1 \right)
=0,\quad
\gamma = \sqrt{ 1+ {\hat P}_1^2 + {\hat P}_2^2}\,, \quad
{{\hat V}_1} = \dfrac{{\hat P}_1}{\gamma}, \quad {{\hat V}_2} = \dfrac{{\hat P}_2}{\gamma},\vspace{1 ex}\\
\dfrac{\partial {\hat P}_1 }{\partial \zt} + {\hat V}_1 \dfrac{\partial {\hat P}_1}{\partial \zr} =
 - {\hat E}_1 - {\hat V}_2 {\hat B}_0,
\quad
\dfrac{\partial {\hat P}_2 }{\partial \zt} + {\hat V}_1 \dfrac{\partial {\hat P}_2}{\partial \zr}=  {\hat V}_1 {\hat B}_0,
\vspace{1 ex}\\
\dfrac{\partial {\hat E}_1 }{\partial \zt} = {\hat N}\, {\hat V}_1.
\end{array}
\label{3gl3}
\end{equation}

From the first and last equations of this system it follows
$$
\dfrac{\partial }{\partial \theta}
\left[ {\hat N} +
\dfrac{\partial }{\partial \rho} {\hat E}_1 \right] = 0.
$$
 Under the traditional assumption of a constant background charge density of stationary ions, this implies a simpler expression for the electron density ${\hat N}(\rho,\theta)$:
\begin{equation}
 {\hat N}(\rho,\theta) = 1 -
\dfrac{\partial  {\hat E}_1(\rho,\theta) }{\partial \rho}.
\label{3gl4}
\end{equation}
Formula (\ref {3gl4}) is a special case of the Gauss theorem \cite {david72}, which in differential dimensional form has the form
$ \Div \bE =  4\,\pi\,e (n - n_0).$
Using  (\ref {3gl4}) in  (\ref {3gl3}), we arrive at the equations, which we will analyze further:
\begin{equation}
\begin{array}{c}
\dfrac{\partial P_1 }{\partial \zt} + V_1 \dfrac{\partial P_1}{\partial \zr} =
 - E_1 - B_0\,V_2,
\quad
\dfrac{\partial P_2 }{\partial \zt} + V_1 \dfrac{\partial P_2}{\partial \zr}= B_0\, V_1,
\vspace{1 ex}\\
\gamma = \sqrt{ 1+ P_1^2 + P_2^2}\,, \quad
V_1 = \dfrac{P_1}{\gamma}, \quad {V_2} = \dfrac{P_2}{\gamma},\vspace{1 ex}\\
\dfrac{\partial E_1 }{\partial \zt} + V_1 \dfrac{\partial E_1}{\partial \zr} =  V_1.
\end{array}
\label{3gl5}
\end{equation}
Here, for the sake of simplicity, we remove the upper symbol from all dimensionless components of the solution.

Consider the initial conditions
\begin{equation}\label{cd1}
     P_1(\rho,0) = P_1^0(\rho), \quad P_2(\rho,0) = P_2^0(\rho), \quad
     E_1(\rho,0) = E^0_1(\rho), \quad
 \rho \in {\mathbb R},
\end{equation}
and we will study in the half-plane $\{(\rho,\theta)\,:\, \rho \in {\mathbb R},\; \theta
> 0\}$ the solution of the Cauchy problem \eqref{3gl5}, \eqref{cd1}.
We  assume that the initial data is at least $C^2$ - smooth.

System  (\ref {3gl5}) is of hyperbolic type. For such systems, there exists, locally in time, a unique solution to the Cauchy problem of the same class as the initial data, in our case it is $ C^2 $. It is also known that for such systems the loss of smoothness by the solution occurs according to one of the following scenarios: either the solution components themselves go to infinity in a finite time, or they remain bounded, but their derivatives \cite {Daf16} turn to infinity. The latter possibility is realized, for example, for homogeneous conservation laws, which include the equations of gas dynamics, where the appearance of a singularity corresponds to the formation of a shock wave.

The article is structured as follows. In Section \ref{S2}, we recall the well-known results concerning the system (\ref {3gl5}) in the special case $ B_0 = V_2 = 0 $, as well as in the non-relativistic case. In Section \ref{S3}, we  find the first integrals of the solution, depending on which we classify the initial data, construct solutions in the form of a traveling wave, and also consider the extended characteristic system. In Section \ref{S4}, we consider the cases of reduction of the extended characteristic system depending on the properties of the first integrals and make changes of variables that allow linearizing it. In Section \ref{S5}, we study small deviations from the equilibrium state, estimate the effect of an external magnetic field on the time of breaking of oscillations, and make a comparison with the nonrelativistic case where possible. In Section \ref{S6}, we  summarize and consider the prospects for further research.

\section{Known results}\label{S2}

{\bf 1.} Let us consider a relativistic analogue of system ~ \eqref {3gl5}, where the speed of particles much less than the speed of light. This leads to the conditions $ \gamma = 1 $ and $ \bp = \bv $, and system ~ \eqref {3gl5} takes the form
\begin{equation}
\begin{array}{c}
\dfrac{\partial V_1 }{\partial \zt} + V_1 \dfrac{\partial V_1}{\partial \zr} =
 - E_1 - B_0\,V_2,
\qquad
\dfrac{\partial V_2 }{\partial \zt} + V_1 \dfrac{\partial V_2}{\partial \zr}= B_0\, V_1,
\vspace{1 ex}\\
\dfrac{\partial E_1 }{\partial \zt} + V_1 \dfrac{\partial E_1}{\partial \zr} =  V_1.
\end{array}
\label{3gl5non}
\end{equation}
This system is much simpler for analysis than ~ \eqref {3gl5}, in \cite {RCharx} a criterion for the formation of singularities in terms of the initial data
\begin{equation}\label{cd2}
     V_1(\rho,0) = V_1^0(\rho), \quad V_2(\rho,0) = V_2^0(\rho), \quad
     E_1(\rho,0) = E_1^0(\rho), \quad
 \rho \in {\mathbb R},
\end{equation}
from the class $C^2$ is obtained.
\begin{theorem} \label{T1}\cite{RCharx} For the existence of a $ C^ 1$ - smooth $ \frac {2 \pi} {\sqrt {1 + B_0^2}} $ - periodic solution $ V_1 (\theta, \rho), \,V_2 (\theta, \rho), \, E_1 (\theta, \rho) $ of the problem  (\ref {3gl5non}), (\ref {cd2}) it is necessary and sufficient that at any point $ \rho \in \mathbb R $ the condition
 \begin{equation} \label {crit2}
\Delta= \left( (V_1^0)' \right) ^ 2 + 2 \, (E_1^0)' +2 B_0\, (V_2^0)' - B_0^2  -1 <0
\end {equation}
holds.
If the opposite  inequality holds at least at one point $ \rho_0 $, then the derivatives of the solution turn to infinity in a finite time.
\end{theorem}

This result says, in particular, that it is easy to construct initial data of a sufficiently general type corresponding to a globally smooth solution in time. For example, if we fix arbitrary initial data and increase $ | B_0 | $, then we are in just such a situation. Thus, the external magnetic field has a regularizing character.

\bigskip

{\bf 2.} Taking relativistic effects into account significantly changes the situation, and the initial data leading to a globally smooth solution must already be selected in a very special way. In the general case, the presence of a relativistic factor acts as a kind of nonlinear resonance, which leads to breaking of oscillations in a certain finite, but possibly quite long time. For the particular case of  system \eqref {3gl5}, corresponding to $ B_0 = P_2 = 0 $, such a problem was solved in \cite {RChZAMP21}. The condition on the initial data, which distinguishes the class of solutions for which a globally smooth solution is possible and, in particular, a traveling wave, looks like
\begin{equation}\label{fi1}
2\sqrt{1+(P_1^0)^2}+(E_1^0)^2\equiv {\rm const}, \quad \rho_0\in \mathbb R.
\end{equation}
In \cite {RChZAMP21} (Theorem 2), a criterion for the formation of singularities in terms of the initial data for this case is obtained.

If the condition \eqref {fi1} is not met, then any small deviation of the initial data from the equilibrium $ P_1 = E_1 = 0 $ leads to the gradient catastrophe.

The aim of this work is to analyze the influence of small perturbations of the initial data for the full system  \eqref {3gl5}. This is a much more complicated problem than the one that was solved for a particular case, which is associated with an increase in the dimension of the phase space of the corresponding characteristic system.

\vspace{1.5em}
\section{Analysis of the characteristic system}\label{S3}

Let us write system \eqref {3gl5} in characteristic form
\begin{eqnarray}\label{char2r}
     \dfrac{dP_1}{d\theta}&=&-E_1-B_0\,V_2,\quad \dfrac{dP_2}{d\theta}=B_0 \,V_1,\quad\dfrac{dE_1}{d\theta}=V_1,\\ \dfrac{d\rho}{d\theta}&=&V_1,\quad V_i=\frac{P_i}{\sqrt{1+P_1^2+P_2^2}},\,i=1,2.\nonumber
     \end{eqnarray}

First integrals of
\ \eqref {char2r} are
\begin{equation}\label{fer}
P_2-B_0 E_1 = K_1, \quad 2\sqrt{1+P_1^2+P_2^2}+E_1^2=K_2.
\end{equation}
Knowledge of the first integrals allows us to express $ E_1 $ and $ P_1 $ in terms of $ P_2 $ and from the second equation \eqref {char2r} obtain an equation for finding $ P_2 (\theta) $ along the characteristic outgoing from the point $\rho_0\in \mathbb R$:
\begin{equation}\label{P2rr}
\dfrac{dP_2}{d\theta}=\pm B_0 \frac{\sqrt{(B_0^2 K_2-(P_2-K_1)^2)^2-4 B_0^4(P_2^2+1)}}{B_0^2 K_2-(P_2-K_1)^2},
\end{equation}
 where $K_i=K_i(\rho_0, 0)$, $i=1,2$. The corresponding constants can be calculated according to \eqref {fer} and, generally speaking, depend on the starting point of the characteristic.
Equation \eqref{P2rr} also allows  to calculate the period of oscillation along a specific characteristic, namely:
\begin{equation*}\label{per_r}
T(\rho_0)= \frac{2}{B_0} \int\limits_{P_2^-}^{P_2^+}  \frac{B_0^2 K_2-(\eta-K_1)^2}{\sqrt{(B_0^2 K_2-(\eta-K_1)^2)^2-4 B_0^4(\eta^2+1)}} \, d\eta,
\end{equation*}
where $P_2^\pm$ are  smaller and larger roots of the equation $$(B_0^2 K_2-(\eta-K_1)^2)^2-4 B_0^4(\eta^2+1)=0.$$
They are chosen such that $P_2(\rho_0,0)\in (P_2^-,P_2^+)$. The specified integral can be expressed in elliptic functions. As in the case of $ B_0 = 0 $, the period of $ T (\rho_0) $ varies from point to point. However, the formula for its determination is much more complicated than the one obtained in \cite {RChZAMP21}, since to do this, you must first express $ P_2 $ in terms of $ P_1 $.

In the case under consideration, traveling waves can also be constructed. To do this, you can use  equation \eqref {P2rr}, fixing in it $K_1 $ and $ K_2 $. If $P_2={\mathcal P}(\xi) $, $\xi=\rho-w \,\theta$, $w=\rm const$, then the equation for the profile  of $ \mathcal P $  is
\begin{equation}\label{P2r}
\dfrac{d\mathcal P}{d\xi}=\pm B_0 \frac{\sqrt{(B_0^2 K_2-({\mathcal P}-K_1)^2)^2-4 B_0^4({\mathcal P}^2+1)}}{-(B_0^2 K_2-({\mathcal P}-K_1)^2)\,w+2 B_0^2{\mathcal P}}.
\end{equation}

We obtain an extended system describing the behavior of the derivatives of the solution along the characteristics:
\begin{equation}\label{char2dr}
     \dfrac{dp_1}{d\theta}=-q_1 p_1-B_0 q_2 -e,\quad \dfrac{dp_2}{d\theta}=-q_1 p_2+B_0 q_1,
     \quad\dfrac{de}{d\theta}=(1-e)q_1,
     \end{equation}
where $q_1=\partial_\rho V_1
$, $q_2=\partial_\rho V_2
$, $p_1=\partial_\rho P_1
$, $p_2=\partial_\rho P_2
$,
$e=\partial_\rho E_1$,
$$q_i=\frac{p_i}{\gamma}-\frac{P_i}{\gamma^3}(p_1 P_1+p_2 P_2), i=1,2, \quad \gamma=\sqrt{1+P_1^2+P_2^2}.$$
System \eqref{char2dr} has the first integral
\begin{equation}\label{C1r}
   p_2=B_0+ C_1(e-1), \quad C_1=\frac{p_2(\rho_0,0)-B_0}{e(\rho_0,0)-1},
     \end{equation}
and can be reduced to two equations
\begin{eqnarray}
  \dfrac{dp_1}{d\theta}= &=& -1-\frac{p_2-B_0}{C_1} -\frac{p_1^2}{\gamma}+p_1 P_1 Q- B_0\left(\frac{p_2}{\gamma}-P_2Q\right),\label{s1}\\
   \dfrac{dp_2}{d\theta}= &=& (B_0-p_2)\left(\frac{p_1}{\gamma}-P_1 Q\right),\quad Q=\frac{P_1 p_1+P_2 p_2}{\gamma^3}.\label{s2}
\end{eqnarray}
If we remember that $ P_1 $ can be expressed in terms of $ P_2 $, and  equation \eqref {P2rr} for  $ P_2 $ is known, then it becomes clear that the analysis of the derivatives of the solution to  problem \eqref {cd1} is reduced to the analysis of an autonomous  system \eqref {s1}, \eqref {s2}, \eqref {P2rr}, and the last equation is decoupled. Thus, formally, the situation is the same as that studied in the relativistic case in the absence of a magnetic field \cite {RChZAMP21}. However, in the presence of an external magnetic field, the system is much more cumbersome.

\bigskip

\section{Cases of reduction and transformation of the system for derivatives}\label{S4}

\subsection{Reduction to one equation}
Let $ p_2 = B_0 $, which corresponds to $ C_1 = 0 $ in the integral \eqref{C1r}.
In this case,  system \eqref {s1}, \eqref {s2} can be reduced to one equation. Indeed, \eqref {s2} holds identically, and \eqref {s1} reduces to
\begin{eqnarray}
  \dfrac{dp_1}{d\theta}= &=& -1 -\frac{B_0^2}{\gamma^3}-\frac{p_1^2}{\gamma^3}-\frac{(p_1 P_2-B_0P_1)^2}{\gamma^3},\nonumber
   \end{eqnarray}
whence it follows that any solution with initial data \eqref {cd1} with $ P_2 ^ 0 = B_0  \rho + \rm const $ loses smoothness in a finite time interval.

\bigskip

\subsection{The case of constant $ K_2 $.}
A more meaningful subclass of solutions are solutions for which the second of   integrals  \eqref {fer}, which coincides with \eqref {fi1} for the particular case $ B_0 = P_2 = 0 $, is identically constant, that is
\begin{equation}\label{K2}
 2 \sqrt {1 +(P_1^0(\rho))^2+ (P_2^0(\rho))^2} +(E_1^0(\rho))^ 2 \equiv K_2={\rm const}, \, \rho\in \mathbb R.
 \end{equation}
For such solutions
\begin{eqnarray} e=-\frac{p_1 P_1+p_2 P_2}{E_1(P_1,P_2)\gamma}, \quad E_1(P_1,P_2)=\pm \sqrt{K_2-2\gamma}.\label{e}\end{eqnarray}
Since  \eqref{C1r} implies  $e=1+\frac{p_2-B_0}{C_1}$ for $C_1\ne 0$, then
there is a relationship between $ p_1 $ and $ p_2 $. If we denote $ s = p_2-B_0 $, then \eqref {s2}  reduces to  system
\begin{eqnarray}
  \dfrac{ds}{d\theta} &=& -L_1 s^2-L_2 s, \label{s21}\\ L_1&=&-\frac{(1+P_2^2)(E_1(P_1,P_2)\gamma+C_1 P_2)}{C_1 \gamma^3 P_1}-\frac{P_1 P_2}{\gamma^3},\nonumber \\
  L_2&=&-\frac{(1+P_2^2)(E_1(P_1,P_2)\gamma+ P_2)}{\gamma^3 P_1}-\frac{B_0 P_1 P_2}{\gamma^3}.
  \nonumber
   \end{eqnarray}
   After replacing $ y = s ^ {- 1} $ the equation \eqref {s21} becomes linear:
   \begin{eqnarray}
  \dfrac{dy}{d\theta} &=& L_1 +L_2 y. \label{s22}
   \end{eqnarray}

\bigskip

\subsection{The case of non-constant $ K_2 $.}
Suppose that the condition \eqref {K2} is not satisfied, and the variables $ e, p_1, p_2 $ are independent. We introduce new variables $u=\frac{e}{p_1}$,  $\lambda=\frac{e-1}{p_1}$ and  $\sigma=\frac{p_2}{p_1}$.
 They satisfy the system
\begin{eqnarray}
  \dfrac{d u}{d\theta} &=& u^2+\frac{B_0\,(1+P_1^2)}{\gamma^3}u\sigma -\frac{B_0\,P_1 P_2}{\gamma^3}u -\frac{ P_1 P_2}{\gamma^3}\sigma+\frac{1+P_2^2}{\gamma^3},\label{ulam1}\\
  \dfrac{d \lambda}{d\theta} &=&\lambda \left(u+ \frac{B_0\,(1+P_1^2)}{\gamma^3}\sigma -\frac{B_0\,P_1 P_2}{\gamma^3}\right),\label{ulam2}
  \\
  \dfrac{d\sigma}{d\theta} &=& \frac{B_0\,(1+P_1^2)}{\gamma^3}\sigma^2 +u\sigma    -\frac{2 B_0\, P_1 P_2}{\gamma^3}\sigma +\frac{B_0\,(1+P_2^2)}{\gamma^3}.\label{ulam3}
\end{eqnarray}

For $ B_0 = P_2 = 0 $ this system decomposes and equations \eqref {ulam1}, \eqref {ulam2} coincide with the corresponding system from \cite{RChZAMP21}.

Notice that
$$
e=\frac{u}{u-\lambda}, \quad p_1=\frac{1}{u-\lambda}, \quad p_2=B_0+C_1 \frac{\lambda}{u-\lambda},\quad \sigma=B_0 u+(C_1-B_0)\lambda,
$$
so $ \sigma $ can be excluded from the system \eqref{ulam1} -- \eqref{ulam3} and we get two equations
\begin{eqnarray}\label{ulam4}
  \dfrac{d u}{d\theta}&=& \left(1+B^2_0\,F_1\right) \,u^2+B_0(C_1-B_0)\,F_1\,u\lambda -2B_0\,F_2 u \nonumber\\&& -(C_1-B_0) F_2\lambda+F_3,\nonumber\\
  \dfrac{d \lambda}{d\theta} &=&\left(1+B^2_0\,F_1\right)\,u\lambda + B_0(C_1-B_0)\,F_1\,\lambda^2 -B_0\,F_2\lambda, 
 \end{eqnarray}
where
\begin{eqnarray}
  F_1=F_1(\theta)=\frac{1+P_1^2}{\gamma^3}, \quad  F_2=F_2(\theta)=\frac{P_1 P_2}{\gamma^3}, \quad
  F_3=F_3(\theta)=\frac{1+P_2^2}{\gamma^3}.\nonumber
 \end{eqnarray}

\subsection{Non-constant $ K_2 $, case of constant $ K_1 $.}
Consider a subclass of solutions determined by the condition $P_2-B_0 E_1 \equiv K_1$ (see \eqref{fer}), for which $C_1=B_0 $,  and the system \eqref {ulam4} takes the form
 \begin{eqnarray}
  \dfrac{d u}{d\theta} &=&M_1 u^2 - 2 M_2 u + M_3, \qquad
  \dfrac{d \lambda}{d\theta} =\lambda (M_1 u -  M_2),
  \label{u}\\
  M_1(\theta)&=&1+
  B^2_0\,F_1\ge 1>0, \quad  M_2(\theta)= B_0\,F_2,\quad  M_3(\theta)=F_3.\nonumber
\end{eqnarray}
The first equation \eqref {u} is then separated from the system. Using standard variable substitutions
\begin{eqnarray}\nonumber
u(\theta)=-\frac{r'(\theta)}{M_1(\theta) r(\theta)}, \quad r(\theta)=z(\theta) \exp \left(-\int\limits_0^\theta \left(M_2(\tau)-\frac{M'_1(\tau)}{2 M_1(\tau)}\right)\, d\tau   \right)
\end{eqnarray}
\eqref {u} reduces to Hill's equation
\begin{eqnarray}\label{Hill}
\dfrac{d^2 z}{d\theta^2}&+&K(\theta) z=0,\\
\label{Kthet}
K(\theta)&=&M_1 M_3 -M_2^2-M_2'-\frac{3}{4}\frac{(M_1')^2}{(M_1)^2}+\frac{M_1''+2 M_2 M_1'}{2 M_1}.
\end{eqnarray}

\subsection{Non-constant $ K_2 $, general case.}

Note that $ \lambda $ does not vanish for finite $ p_1 $, and we make one more change of variables: $q_1=\frac{u}{\lambda}$, $q_2=\lambda^{-1}$. With respect to the new variables, we obtain a linear inhomogeneous system of equations
\begin{eqnarray}
  \dfrac{d q_1}{d\theta} &=& - B_0 F_2 \, q_1 + F_3 \, q_2-(C_1- B_0) F_2, \label{q1q2_1}\\
   \dfrac{d q_2}{d\theta} &=& -(1+B_0^2 F_1) \, q_1 + B_0 F_2 \, q_2-B_0(C_1- B_0) F_1.
  \label{q1q2_2}
 \end{eqnarray}
 This system can be reduced to a linear inhomogeneous second-order equation for $q_1(\theta)$:
 \begin{eqnarray*}
  \dfrac{d^2 q_1}{d\theta^2}&+&G_3(\theta)  \dfrac{d q_1}{d\theta}+G_2(\theta) q_1 + G_1(\theta)=0,\\
  G_1&=&(C_1-B_0) \left(F'_2-B_0 F_2^2+B_0 F_1 F_3- \frac{F_2 F_3'}{F_3}\right),\\
 G_2&=&F_3(1+B_0^2 F_1)- F_2^2 B_0^2+ B_0 F_2'-\frac{ B_0 F_2 F_3'}{F_3},\\
  G_3&=&-\frac{F_3'}{F_3},
  \label{G}
 \end{eqnarray*}
whence, in turn, using the standard change of variables
 \begin{eqnarray} \label{q1w}
 q_1(\theta)= e^{-\frac12\,\int\limits_0^\theta G_3(\zeta) d\zeta}\,  w(\theta)= \frac{w(\theta) \sqrt{F_3(\theta)}}{\sqrt{F_3(0)}}
 \end{eqnarray}
the term with the first derivative can be excluded:
 \begin{eqnarray} \label{N}
  \dfrac{d^2 w}{d\theta^2}+N_2(\theta) w + N_1(\theta)=0,\\
  N_1=G_1 e^{\frac12\,\int\limits_0^\theta G_3(\zeta) d\zeta}=\frac{G_1 \sqrt{F_3(0)}}{\sqrt{F_3}},\quad
  N_2=G_2-\frac14 G_3^2-\frac12 G_3'.\nonumber
 \end{eqnarray}

It is easy to check that for constant $ K_1 $, when the equality $ C_1 = B_0 $ holds, the equation \eqref {N} is homogeneous and coincides with \eqref {Hill}.

Since $ q_1 = \dfrac {e} {e-1} $, in order to determine whether $ e $ goes to infinity (together  with  $ p_1 $ and $ p_2 $), it is enough to determine whether there is a time  $ \theta_* $ such that $ q_1 (\theta_*) = 1 $. However, it is impossible to obtain an explicit form of the solution of system \eqref {q1q2_1}, \eqref {q1q2_2}, despite the fact that the dependence of the functions $ F_1, \, F_2, \, F_3 $ on time can be considered known. It is for this reason that we have to confine ourselves to the study of small deviations from the equilibrium state.


\section{Study of small deviations from the equilibrium state}\label{S5}
In \cite {RChZAMP21}, we proved that in the general situation a solution corresponding to an arbitrarily small deviation from the zero equilibrium state, in the relativistic case, necessarily loses its smoothness. Below  investigate whether this result remains valid in the presence of an external magnetic field.



\subsection{Solutions with non-constant $ K_2 $}

\begin{theorem}\label{T2}
Let the initial data \eqref {cd1} be such that
$$ 2 \sqrt {1 +(P_1^0(\rho))^2+ (P_2^0(\rho))^2} +(E_1^0(\rho))^ 2 $$ does not equal identically to a constant. Then
  derivatives of any solution to the Cauchy problem \eqref {3gl5}, \eqref {cd1} corresponding to an arbitrarily small deviation of the initial data \eqref {cd1}
from the state $ P_1 = P_2 = E_1 = 0 $ turn to infinity in a finite time.
\end{theorem}

\proof
1. Let us prove the theorem first for the case  $P_2^0(\rho)-B_0 E_1^0(\rho) = K_1, \, \rho\in \mathbb R,$
when the equation \eqref {N} is homogeneous and coincides with \eqref {Hill}. 

 We modify the method used in \cite {RChZAMP21} and show that any solution of equation \eqref {Hill}  corresponding to a small deviation from the equilibrium position is oscillating with increasing amplitude. To do this, we have to do cumbersome, but standard calculations. To distinguish a class of small perturbations, we assume that $K_2=2+\epsilon^2, $ $K_1=0$,   $\epsilon\ll 1$. Then, leaving on the right-hand side of \eqref {P2rr} terms of order at most two, we obtain
\begin{equation*}
P_2(\theta)= \epsilon \frac{B_0}{\sqrt{1+B_0^2}}\,\sin(\sqrt{1+B_0^2} \theta)+O(\epsilon^2),\label{P2}
\end{equation*}
 where we set the phase to be zero without loss of generality. The $ P_1 (\theta) $ function is found from the condition
 \begin{equation*}
 P^2_1=\frac14 \left(2+\epsilon^2 -\frac{P_2^2}{B_0^2}\right)^2-P_2^2-1.
 \label{P1}
\end{equation*}
After substituting $ P_1 $ and $ P_2 $ into \eqref {Kthet} and expanding this expression in a series in $ \epsilon $ up to a power not higher than two, we obtain
\begin{eqnarray}\label{K_series}
K(\theta)=\hat a-2\hat b \cos (2\sqrt{1+B_0^2}\theta )&+& O(\epsilon^4),\\
\hat a=(1+B_0^2)-\frac{8 B_0^2(1+B_0^2)+3}{4(1+B_0^2)}\epsilon^2,&\quad &\hat b= \frac{8 B_0^2(1+B_0^2)+3}{8(1+B_0^2)}\epsilon^2.\nonumber
\end{eqnarray}
Let us make the change $\tau=\sqrt{1+B_0^2} \theta$.
 If we neglect the terms higher than the second order in \eqref {K_series} and substitute the result in \eqref {Hill}, we get the well-studied Mathieu equation (e.g.,\cite {BEr67}, Chapter 16)
 \begin{eqnarray}\label{Matier}
\dfrac{d^2 z}{d \tau^2}+(a-2 b \cos 2 \tau) z=0,\quad
a=\frac {\hat a}{1+B_0^2},\quad b=\frac {\hat b}{1+B_0^2}.
\end{eqnarray}
 According to the Floquet theory, the boundedness or unboundedness of the solution of such an equation, together with its derivative, is completely determined by its characteristic exponent $ \mu $, defined as the solution to the equation $\cosh \mu
\pi = z_1(\pi),$ where $z_1(\theta)$ is the solution of the Mathieu equation with initial conditions
 $z_1(0)=1$, $z_1'(0)=0$. Unboundedness takes place for real $ \mu $, that is, if $|\cosh \mu \pi|>1$. According to the asymptotic formula \cite {BEr67}, Section 16.3 (2), which (taking into account the misprint in the sign in the formula from Section 16.2 (15)) has the form
 $$\cosh \mu \pi = \cos
\sqrt a \pi-\frac{\pi b^2}{(1-a)\sqrt a }\sin \sqrt a \pi
+O(b^4),\,b\to 0,$$ we get
\begin{equation}\label{K1B0}
\cosh \mu
\pi=-1-\frac{\pi^2}{2048}\frac{(8 B_0^2(1+B_0^2)+3)^3}{(1+B_0^2)^6}\epsilon^6+O(\epsilon^8)<-1.
\end{equation}
 We see that  $ K(\zt) > 0 $ (see \eqref{Hill}, \eqref{Kthet}), therefore, any solution to the Mathieu equation oscillates.

  Since $ u = -\frac{r '}{M_1 r} $, then according to the second equation \eqref{Hill},
  $ \frac{\lambda '}{\lambda} = - \frac{r '}{r}-M_2 $, that is $\lambda r =\lambda(0) e^{-\int\limits_0^\theta M_2(\zeta) d\zeta}$.

We set without loss of generality $ z (0) = r(0)=1 $, then $$ r '(0) = - M_1(0),\quad  z'(0)= - M_1(0)+M_2(0)-\frac{M_1'(0)}{2M_1(0)}.$$ Further,
$$
\dfrac{u}{\lambda} = \dfrac{e}{e-1} = -\,\dfrac{r'}{M_1 r \lambda}=-\frac{r'}{M_1\lambda(0) e^{-\int\limits_0^\theta M_2(\zeta) d\zeta}}
=\frac{-z'+(M_2-\frac{M_1'}{2M_1})z}{\sqrt{M_1}\lambda(0)}.
$$
From these considerations it is clear that if $ e $ at some point $ \theta = \theta_* $ becomes infinity, then at this point
$ \frac {-z '+ (M_2- \frac {M_1'} {2M_1}) z} {\sqrt {M_1} \lambda (0)} $ becomes $ 1 $. If there is no such moment, then $ e $, and with it $ p_1 $ and $ p_2 $, remain bounded.

However, as we have shown, any solution of \eqref {Matier} oscillates and the amplitude of its oscillations increases. Therefore, any linear combination of a solution with its derivative has the same property (note that the expression $ M_2- \ frac {M_1 '} {2M_1} $ for all $ \theta $ remains bounded). Therefore, $ \dfrac {u} {\lambda} $ at some point in time will reach any predetermined constant, including one.

So, we have proved the theorem for the case of constant $ K_1 $.
\bigskip

2. Let us consider the general case of non-constant $ K_1 $.
The first terms of the expansion of inhomogeneity of equation \eqref {N} are of the form
\begin{equation*}
N_1(\theta)=B_0 (B_0-C_1)\left(1+\frac{5+12 B_0^2}{8(1+B_0^2)}\epsilon^2- \frac{3+10 B_0^2}{8(1+B_0^2)}\epsilon^2 \,  \cos (2\sqrt{1+B_0^2}\theta ) \right)+ O(\epsilon^4).
\end{equation*}
This is a bounded continuous function for $ \theta> 0 $. We denote by $ \bar w (\theta) $ a fixed particular solution of the equation \eqref {N}, for example, with the initial conditions $ \bar w (0) = \bar w '(0) = 0 $. It is easy to see that in the zero approximation this is a bounded function. Approximations of the following orders are found from linear second-order equations with constant coefficients and do not contain resonance terms, that is, they are also bounded. The general solution \eqref {N} consists of the general solution of the corresponding homogeneous equation, which, as we proved above, is oscillating with an exponentially growing amplitude, and $ \bar w (\theta) $. Thus, for any initial conditions, $ w (\theta) $ in a finite time will reach any predetermined constant in a finite time. Therefore, $ q_1 $, which, according to \eqref {q1w}, is obtained by multiplying $ w $ by a bounded function, will reach the value 1 in a finite time.

 Thus, Theorem 2 is completely proved.


\subsubsection{Influence of $ B_0 $ on the breaking time.}
As follows from Theorem 2, in the nonrelativistic case, an increase in $ | B_0 | $ always extends the class of smooth initial data, except for the case of constancy of the integral $ K_1 $ for all characteristics. In this exceptional case, $ B_0 $ participates in the selection of the initial data and its increase has the opposite effect.

In the relativistic case, in the situation where the integral $ K_1 $ is constant, a similar phenomenon also arises of a decrease in the lifetime of a smooth solution with increasing $ | B_0 | $. Indeed, as follows from \eqref {K1B0},
the correction to the quantity responsible for the exponential growth of the oscillation amplitude has the form
$$-\frac{\pi^2}{2048}\frac{(8 B_0^2(1+B_0^2)+3)^3}{(1+B_0^2)^6}\epsilon^6+o(\epsilon^6).$$
As $ | B_0 | $ increases from zero, this correction does increase in absolute value, but it has a limit at $ | B_0 | \to \infty $, which indicates that the dependence of the time of breaking of oscillations on the strength of the magnetic field ceases to be significant at large $ | B_0 | $.

When $ K_1 $ is not constant, the conclusions are not so unambiguous. The problem is that here it is necessary to take into account the influence of the particular solution of  equation \eqref {N}, which is also oscillating and the oscillations can be mutually canceled out.

In addition, it should be noted that the exponential growth of the oscillation amplitude of the solution to the homogeneous equation manifests itself in a significantly larger order of smallness in $ \epsilon $, therefore, at sufficiently small times, the processes associated with the linearized equation, that is, with terms of order zero in $ \epsilon $. But this means that we are in the framework of the nonrelativistic model, for which, in the general case, the increase in $ | B_0 | $ has a regularizing character.

\bigskip

\subsection{Solutions with constant $ K_2$.}
It is clear that in this case not all solutions lose smoothness in a finite time. Indeed, a traveling wave solution \eqref {P2r} falls into the class of solutions with constant $ K_2 $. Such a solution can be constructed in the form of a small deviation from the zero equilibrium position, and its derivatives do not vanish into infinity.

\begin{theorem}\label{T4}
Let the initial data \eqref {cd1} be such that  $$ 2 \sqrt {1 +(P_1^0(\rho))^2+ (P_2^0(\rho))^2} +(E_1^0(\rho))^ 2 \equiv K_2={\rm const}, \, \rho\in \mathbb R.$$ Then there is $ \epsilon_0> 0 $ such that
   the solution of the Cauchy problem \eqref {3gl5}, \eqref {cd1} corresponding to a perturbation of the initial data \eqref {cd1} from the state $ P_1 = P_2 = E_1 = 0 $ of order $ \epsilon <\epsilon_0 $ keeps $ C^1 $ - smoothness if and only if at the initial moment of time the condition
\begin{equation}\label{K2c}
2e+2B_0 p_2-2B^2_0-1<0
\end{equation}
holds.
\end{theorem}

\bigskip

For the {\it proof}, as in the previous subsection, we put  $K_2=2+\epsilon^2, $ $K_1=0$,   $\epsilon\ll 1$, and get
$$
\begin{array}{c}
P_2(\theta)=  \frac{\epsilon B_0}{\sqrt{1+B_0^2}}\,\sin(\sqrt{1+B_0^2} \theta)+O(\epsilon^2),\\  P_1(\theta)= \epsilon \,\cos(\sqrt{1+B_0^2} \theta)+O(\epsilon^2),\\
 E_1(\theta)= \frac{\epsilon}{\sqrt{1+B_0^2}}\,\sin(\sqrt{1+B_0^2} \theta)+O(\epsilon^2),\quad \epsilon\to 0.
\end{array}
$$
We use  equation \eqref {s22}. Its solution has the form
$$
y(\theta)= \left(y(0)+B_0+\frac{1}{C_1}\right)\,\left(\cos(\sqrt{1+B_0^2}\theta)\right) ^{\frac{1}{1+B_0^2}}-\left(B_0+\frac{1}{C_1}\right)+O(\epsilon).
$$
It is easy to calculate that the zero approximation $ y (\theta) $ does not vanish, that is, $ p_2 $, together with the rest of the derivatives, does not vanish
provided that at the initial moment  inequality \eqref{K2c} holds.


\subsubsection{Comparison with the nonrelativistic case.} Let us compare this result with Theorem 1.

Considering that $ P_2 (0) = E_1 (0) $ according to the assumption that $ K_1 = 0 $ and $ p_1 (0) = 0 $
(the value of $ p_1 $ can be found from \eqref {e}), then \eqref{K2c} can be compared with the necessary and sufficient condition for the preservation of smoothness by the solution in the nonrelativistic case \eqref {crit2}, which in this situation has the form $$ 2e + 2B_0 p_2- B ^ 2_0-1 <0. $$ This comparison shows how much more stringent condition guarantees the non-overturning of sufficiently small oscillations in the relativistic case. In addition, we note that the presence of a magnetic field expands the class of initial data corresponding to globally  in time smooth solutions.


\section{Conclusion}\label{S6}

We have investigated the simplest model of relativistic oscillations of a cold plasma in a constant magnetic field. It was found that even small perturbations of the trivial state of rest, as a rule, eventually lead to the formation of singularities of the solution (breaking of oscillations). However, at short times, the nature of the solution is largely determined by the leading terms of the expansion in a small parameter, which allows us to conclude that the external magnetic field has a regularizing effect. However, if in the nonrelativistic case the breaking can be completely eliminated by means of an external magnetic field, but in the relativistic case it can only be delayed.

The success of the study in this case is determined by the fact that the system of partial differential equations is nonstrictly hyperbolic, which makes it possible to write an extended system along one characteristic direction and thereby reduce it to a problem for ordinary differential equations. For the complete, unreduced, cold plasma model, this property no longer exists. However, there is every reason to believe that the technique developed in this paper can help in "splitting into processes" for the full model, where the process of breaking of oscillations is accompanied by their wave transfer.

\section*{Acknowledgments}
Supported by the Ministry of Education and Science of the Russian Federation as part of the program of the Moscow Center for Fundamental and Applied Mathematics under the agreement 075-15-2019-1621.

The authors are grateful to E.V.Chizhonkov for  stimulating discussions and constant interest.


\end{document}